\documentclass[aps,prl,twocolumn,showpacs,preprintnumbers,
amsmath,amssymb,superscriptaddress]{revtex4}

\usepackage{graphicx} \usepackage{color} \usepackage{dcolumn}
\usepackage{bm}

\definecolor{violet}{rgb}{0.5,0,0.5}
\makeatletter \def\@dotsep{5} \makeatother

\newcommand{\tdp}{{\hbox{$\pmb{\pmb{\boldsymbol{\blacktriangledown}}}
 \mkern-20mu$\raise0.15ex\hbox{\textbf{- -}}}}}

\newcommand{\tud}{{\hbox{$\pmb{\pmb{\boldsymbol{\bigtriangleup}}}$}}}
\begin{document}

\title{Evidence for Macroscopic Quantum Tunneling of Phase Slips in Long One-Dimensional Superconducting Al Wires}

\author{Fabio Altomare} \author{Albert M. Chang} \affiliation{Physics
  Department, Duke University, Durham, NC 27708}\altaffiliation[Current address.]{} \affiliation{Physics
  Department, Purdue University, West Lafayette, IN 47906 }

\author{Michael R. Melloch} \affiliation{School of Electrical and
  Computer Engineering, Purdue University, West Lafayette, IN 47906}

\author{Yuguang Hong} \author{Charles W. Tu} \affiliation{Department
  of Computer and Electrical Engineering, UCSD, La Jolla, CA 92093}

\date{\today}

\begin{abstract}
Quantum phase slips have received much attention due to their 
relevance to superfluids in reduced dimensions  and to models 
of cosmic string production in the Early Universe.
Their establishment in one-dimensional superconductors has remained
 controversial. Here we study the nonlinear voltage-current characteristics and
 linear resistance  in  long superconducting Al wires with lateral 
dimensions $\sim$ 5 nm.
 We find that, in a magnetic field and at temperatures well below 
the superconducting transition, the observed behaviors can be described by the
  non-classical, macroscopic quantum tunneling of phase slips, and
  are inconsistent with the thermal-activation of phase slips.


\end{abstract}


\maketitle

What is the resistance of a superconductor below its superconducting
transition temperature? For a three-dimensional superconductor, the answer is
the obvious one -- zero. But in one-dimension (1D), by which we mean very
thin very long wires, quantum fluctuations destroy the zero resistance state.
Phase slips -- small superconducting regions that become normal,
allowing the phase of the order parameter to rapidly change by 
$2\pi$ -- give rise to residual resistance and can even quench superconductivity
completely. The tiny cross-section of 1D nanowires reduces the free energy
barrier arising from a loss of condensation energy 
for the creation of phase slips. 
Thermal activation of phase slips (TAPS) across this barrier 
is responsible for the residual resistance just below $T_c$ 
(LAMH picture) \cite{la}. On the other hand,  macroscopic quantum
tunneling of phase slips (QTPS) 
as the source of residual resistance
 at low temperatures has remained
controversial despite intense experimental effort 
\cite{Gio88,Duan,Bezryadin, Lmb+01,tian,ion-beam-sputtering-aluminum,
rogachev}. 
 The observation of macroscopic quantum tunneling is of significance not 
 only for 1D superconductivity, but also for understanding 
the decoherence of quantum systems
 due to interaction with their environment.
Here we clearly establish the presence of QTPS in 
superconducting (SC)
Al nanowires, showing the 
dramatic effect they have on the voltage-current (V-I) 
characteristics of the wire.

Phase slips in 1D SC nanowires have
traditionally been  studied by
measuring the linear resistance as a function of the temperature 
through the normal-superconducting (N-S) transition. Any resistance in excess 
to that predicted by thermal activation is attributed to 
macroscopic tunneling of phase 
slips \cite{Gio88,Duan,Bezryadin,Lmb+01,tian,ion-beam-sputtering-aluminum}.
However, this attribution can be flawed because of weak links in the wire   
resulting from inhomogeneities \cite{Duan,ion-beam-sputtering-aluminum}.  
Moreover, the fitting 
of the excess residual resistance to quantum-phase-slip 
expressions
\cite{Gio88,Lmb+01,ion-beam-sputtering-aluminum,tian} often
necessitated an ad hoc reduction in the free-energy barrier.
Recently Rogachev \textit{et al.} \cite{rogachev} examined the nonlinear
V-I  dependence of $Mo_{0.79}Ge_{0.21}$
SC nanowires: 
they  found that the deduced residual
resistance, spanning 11 orders in magnitude, followed the prediction of 
classical TAPS  alone, thus contradicting previous claims of QTPS
behavior based on residual linear resistance by the same 
group \cite{Bezryadin,Lmb+01}.

In this Letter, we examine the nonlinear V-I characteristics 
and linear resistance of  long aluminum nanowires, the narrowest one 
with dimension 5.2 nm x 6.1 nm x 100 $\mu$m (21 x 24 x 400,000 atoms).  
Our wires  are much longer than those  of similar cross section reported in 
the literature ($\sim 0.5 \mu$m long)\cite{Bezryadin,Lmb+01, rogachev} and
 the ratio of the low temperature SC coherence length to the width (or height) 
is also much larger, thus placing us in a regime 
distinct from previous works. Specifically, we study the residual resistance deduced from the nonlinear V-I
characteristics in the superconducting  state below the
 critical current. Our main
finding is that the V-I dependence and residual resistance are
inconsistent with the classical LAMH behavior, but instead are well
described by quantum expressions, either derived  from an extension of
the classical model \cite{Gio88} or in a recently proposed
power-law form \cite{Gz01,khlebnikov,khleb_epaps}.  The good fits 
to the different
quantum expressions, which closely overlap each other, are
corroborated by measured residual linear resistance, achieving full
consistency within the quantum scenario using a single set of fitting
parameters.   
 Our results demonstrate the importance of non-classical,
 quantum phase slips in ultranarrow, 1D superconducting aluminum wires.

The nanowires were fabricated by thermally evaporating 
aluminum onto a narrow, 8 nm-wide, MBE (molecular-beam-epitaxy) grown InP ridge, while
at once linking and partially covering the four-terminal Au/Ti
measurement pads 
(Fig. 1(b)).
 The fabrication is  described elsewhere\cite{altomare_apl}.  Magnetoresistance at 4.2 K (above $T_c$) was
used to characterize the wires  (Table 1). 
The calculated dirty limit $\xi$ \cite{TinkhamBook}
 ($\approx$ 94 nm for s1,  $\approx$ 128 nm for s2)
 far exceeds the lateral dimensions ($\sim 5-9$~nm).  

\begin{table}
  \begin{tabular}{ccccccccc}\hline
    & $L$ & $w$& $t$& $R_N$ & $\rho$& $l_e$ & $\xi$ & $J_c$ (0.35 K)\\ 
    & $\mu$m & nm & nm & $k\Omega$ & $\mu \Omega cm$ & nm & nm & $10^5
    A/cm^2$\\
    \hline s1 & 10 & 6.9 &9.0 & 8.3 & 5.1 & 7.6 & 94 & 12.1 \\
    s2&100&5.25 &6.09 &86 &2.8& 14.3&128 &13.1 \\
    \hline
  \end{tabular}
  \caption{Parameters for sample s1 and s2, including $L$ (length),
    $w$ (width), $t$ (thickness), $R_N$ (normal state resistance), $\rho$
    (resistivity), $l_e$ (mean free path), $\xi$ (GL coherence length),
    and $J_c$ (critical current density).
    $w$, $t$, $\rho$ and $l_e$ were deduced based on magnetoresistance
    measurements \cite{altomare_apl}. 
 $\xi$ was calculated using 
    $\xi \!=\!0.85\sqrt{\xi _0
      l_e}$ where $\xi_0 \!=\!1600~nm$ is the GL superconducting coherence
    length in bulk aluminum.}
  \label{tab:tab1}
\end{table}

\begin{figure}
 \includegraphics[clip,width=3.1in]{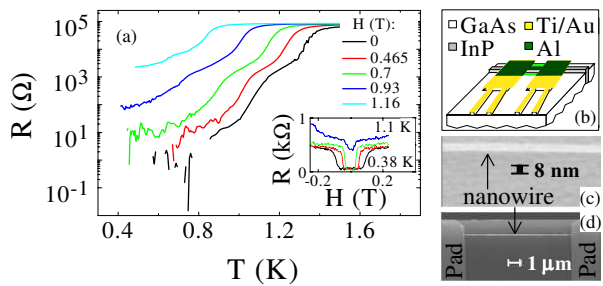}
  \caption{(Color)
    (a) Linear resistance versus temperature for wire s2
    in different magnetic fields;
    the pad series resistance has been subtracted\cite{prl_7}.
    Inset: The pad series resistance increases 
    at magnetic fields beyond the critical field of the pad regions
    covered by aluminum 
    (T=0.38, 0.80, 0.99, 1.10 K). 
    (b)-(d) Schematic and SEM images of a nominally 8 nm
    wide Al nanowire (similar to samples s1 and s2): the Al layer does
    not entirely cover the Au/Ti pads which were measured in series
    with the superconducting wire. 
}
  \label{fig:fig1}
\end{figure}

To investigate the SC behavior, the linear resistance through the N-S
transition was measured at a current of 1.6 nA (Fig. 1(a)), 
while the nonlinear V-I characteristics was measured in both the constant-current 
 (Figs. 2(a)-(b) and 3) and constant-voltage modes 
 below $T_c$ \cite{prl_6}.  The linear resistance
drops by several orders of magnitude from the normal state 
resistance ($R_N$) through $T_c$, but remains finite.  
In Fig. 1(a) we show the linear 
resistance versus temperature for wire s2 in different magnetic fields ($H$)\cite{prl_7}. 
At a given temperature, the resistance 
increases with increasing $H$ as superconductivity weakens.  
In the constant-current mode, finite residual nonlinear voltage is observed 
in the V-I curves below the critical current jump.  Such residual voltage is unobservable 
in large wires. 
In the constant-voltage mode, the V-I curves 
exhibit non-hysteretic voltage steps and S-shaped curves down to $T/T_c
\approx 0.2$, typical of 1D superconductors \cite{tian,vodolazov}.  
Evidences for wire homogeneity include: 1) comparable
critical current density in
wires s1 and s2 
2) all V-I traces
in the constant-current mode showed a single critical current
jump. 

To assess the importance of QTPS, we focus on 
the residual resistance and residual voltage below $T_c$ 
for sample s2; the wider s1 will be  used for control as discussed below.
The main datasets consist of the nonlinear V-I curves in constant-current mode
(Figs. 2(a)-(b) and 3 (black curves)), and the 
linear resistance versus $T$.  
This latter resistance is obtained  in two ways: 
(i) from the 
data in Fig. 1(a) 
\cite{prl_7}  
(Figs. 2(c)-(d) (black curves)), 
and (ii) from the fits to the nonlinear V-I curves according to 
Eqs. (1b) and (3) below 
($\pmb{\pmb{\boldsymbol{\bigtriangleup}}}$, Figs. 2(e)-(f)).  It will be demonstrated through 
quantitative analysis that this entire set of data is inconsistent with the 
TAPS scenario, as evidenced by the poor fits to the LAMH expressions 
(Eqs. (1a) and (1b)) shown in Figs. 2 and  3 (blue curves). 
Instead  the inclusion of the QTPS contributions 
(Giordano (GIO) model Eqs. (2) and (3)) in the fits
is essential in providing a 
fully consistent description 
based on a single set of parameters (see red curves in Figs. 2 and 3). 

\begin{figure}
  \includegraphics[clip,width=3.2 in]{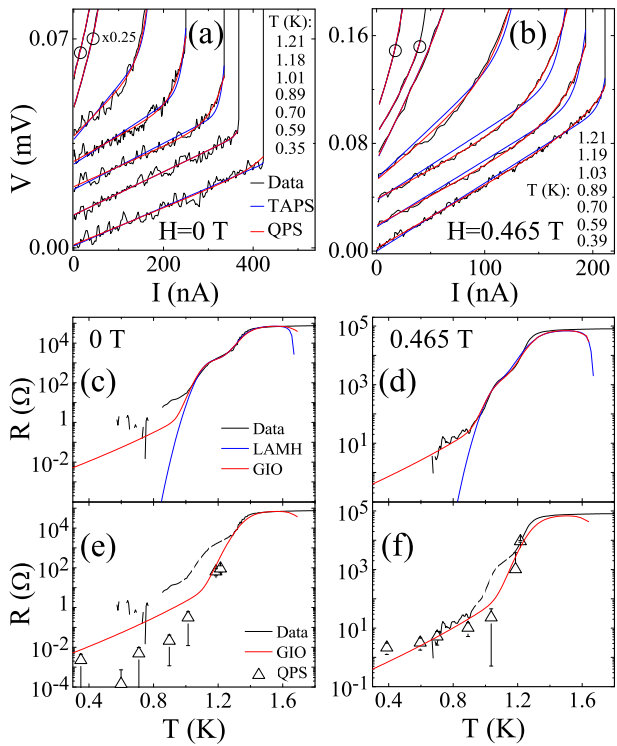}
\caption{(Color) Nonlinear V-I curves and linear resistance 
  for sample s2 at different magnetic fields ($H$):
  Black curves--data; red curves--fits to the GIO ($\equiv$GIO + LAMH)
expressions
  for QPS ($\equiv$TAPS + QTPS), and blue curves--fits to the LAMH 
  expressions for TAPS  alone.
  (a)-(b) V-I curves offset for clarity.
   The fits to QPS are of higher quality compared to TAPS; each fit includes
 a series resistance term $V_S$.
  These V-I curves can be
  fitted equally well by a power law form $V=V_S +K\cdot
  (I/I_k)^{\nu}$ where $12 \gtrsim \nu \gtrsim 3.2$ for $0 \leq H \leq
  1.05$~T\cite{khlebnikov, khleb_epaps}.
  (c)-(d) Linear resistance after background subtraction (see Fig.~1(a)
  and \cite{prl_7}).  The LAMH fits are poor at low $T$.
(e)-(f) The resistance contribution due to phase slips ($R_{QPS}$)
extracted in \textit{(a)-(b)} from fits to the V-I curves using the GIO expressions (discrete points, $\tud$). 
 $R_{QPS}$ and the linear resistance   from \textit{(c)-(d)} are
 refitted using the GIO expressions (red) with
  the same $a_{GIO}(=1.2)$
while disregarding the irrelevant shoulder feature (dashed line).
}
\label{fig:fig3}
\end{figure}

Thermal activation of phase slips, important at $T\lesssim T_c$, is well
described by the expressions derived by Langer Ambegaokar, McCumber,
Halperin (LAMH)\cite{la,prl_2}:
\begin{subequations}
\begin{eqnarray}
     &  R_{LAMH}\!=\!R_q\displaystyle{\frac{\Omega}{k_BT}}\exp{\left(\displaystyle{-\frac{\Delta F}{k_BT}}\right)},\\\label{eq:La}
     & V_{TAPS}\!=\!I_0R_{LAMH}\sinh{\left(I/I_0\right)}, \label{eq:L1}
\end{eqnarray}
\end{subequations}

\noindent where the quantum resistance $R_q\!=\!\pi\hbar/2e^2$,
attempt frequency $\Omega\!=\!(L/\xi)(\Delta F/k_BT)^{1/2}(\hbar/\tau
_{\scriptscriptstyle GL})$, free energy barrier $\Delta
F\!=\!(8\sqrt{2}/3)(H_{th}^2/8\pi)A\xi$, and Ginzburg-Landau relaxation
time $\tau _{\scriptscriptstyle GL}\!=\!(\pi/8)[\hbar/k_B(T-T_c)]$,
$I_0\!=\!4ek_BT/h$.  $L$ is the wire length, $T$ the temperature,
$k_B$ the Boltzmann constant, $H_{th}$ the thermodynamical critical
field, $A$ the wire cross section, and $\xi$ the GL coherence length.

Quantum tunneling of phase slips, which takes place in parallel with thermal
 activation,
 is expected to dominate at low $T$.
 Giordano\cite{Gio88} proposed a quantum
form by replacing $kT$ with $(1/a_{GIO})(\hbar/\tau_{GL})$\cite{prl_2,GIO}:
\begin{equation}
  \label{eq:GIO}
  R_{GIO}\!=R_q\frac{L}{\xi}\sqrt{a_{GIO}\frac{\Delta F}{\hbar/\tau
      _{\scriptscriptstyle GL}}}\exp{\left(-a_{GIO}\frac{\Delta
        F}{\hbar/\tau _{\scriptscriptstyle GL}}\right)}
\end{equation}
\noindent where $a_{GIO}$ is a numerical constant of order unity and,
 in analogy with the thermal case, we propose that\cite{prl_2,GIO}: 
\begin{equation}
  \label{eq:non_linear_IV}
  V_{QTPS}\!=\!I_{GIO}R_{GIO}\sinh{\left(I/I_{GIO}\right)}
\end{equation}
where $I_{GIO}\!=\!2e/(\pi\tau _{\scriptscriptstyle GL} a_{GIO})$. 
In this quantum
case, a resistance similar to the Giordano expression (and numerically
equivalent) has been recently derived, on a microscopic basis, by
Golubev and Zaikin \cite{Gz01} and by Khlebnikov and Pryadko
\cite{khlebnikov}.  In the nonlinear regime, both theories predict a
crossover from an exponential dependence to a power-law behavior
($V\propto I^{\nu}$) at low $T$.

To begin the analysis, we first extract the parameters, 
$T_c$ and $\xi$ at each magnetic field and a single $a_{GIO}$, 
by fitting to the linear $R$ versus $T$ traces in Figs. 2(c)-(d);
$T_c$ and $a_{GIO}$ are input parameters in the analysis of the V-I curves.
Below $T_c$, a wire is modeled as a normal
and a superconducting wire in parallel, 
the latter with a resistance produced by the relevant phase slips mechanism\cite{Lmb+01}.
To account for the shoulder feature at a resistance $\approx 1/20$ of $R_N$, 
we assume our wire to be composed of segments of two slightly different cross
sectional areas, with a fixed ratio of 90\% ($\Delta F\! \propto
\!A$). 
The thinner segments corresponding to the smaller cross sectional area 
have a larger $T_c$\cite{ion-beam-sputtering-aluminum}.
These segments, with critical temperature (resistance) $T_{c1}$
($R_1$), are responsible for $\gtrsim{}$95\% of $R_N$.  At zero field, the
fitting parameters are $T_{c1}(H\!=\!0)$, $T_{c2}(H\!=\!0)$,
$\xi(H\!=\!0)$, $R_1$, and $a_{GIO}$, while at $H\geq 0.465$~T, only
$T_{c1}(H)$, $T_{c2}(H)$, $\xi(H)$ are varied.
Fitting  using two $\xi$s produced unphysical $\xi$
dependence on $H$ and was discarded.
The fits to the GIO (red)  and LAMH (blue) theory 
are presented in Figs. 2(c)-(d).  At
$H=0$ T both fits reproduced the data over several decades in
resistance and are nearly equivalent above $\approx$5 $\Omega$. 
At higher field the GIO theory better models the data, 
particularly at low $T$\cite{reb_3_1}.
The fitting parameters $T_{c1}$ and $\xi$ ($\pmb{\pmb{\boldsymbol{\circ}}}$) 
are presented in Fig. 4 (a)-(b).  Their $H$ dependence
can be fitted to simple theoretical expressions as shown.
The single value of $a_{GIO}\!=\!1.2$ we obtain, of order unity as expected, 
will be used throughout  all the subsequent analysis.

\begin{figure}
  \centering
  \includegraphics[clip,width=3.3 in]{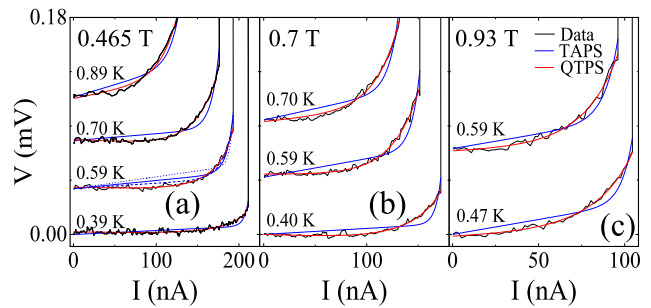}

\caption{(Color) The nonlinear V-I curves, at $T/T_c\lesssim 0.5$ are refitted with 
  $V=V_{QTPS}+V_S$ and $V=V_{TAPS}+V_S$ (see text).(a)-(c) The  
data (black), QTPS (red) and TAPS (blue) fits and
are shown after subtracting the linear background.
Dotted (dashed) line in 3(a) is the fit to the TAPS expression 
   with a 50\% reduction (increase) in temperature, respectively.
}  \label{fig:fig4}
\end{figure}
Despite the success in fitting to the linear resistance, it is
not possible to completely rule out weak
links as the source of residual resistance.  To clearly demonstrate the
importance of quantum phase slips, it is necessary to analyze the
nonlinear V-I dependence.  Generalizing the nonlinear analysis of Rogachev
et al.\cite{rogachev} by including QTPS while also accounting for series
resistances, 
the total voltage drop across the
superconducting nanowire is $V\!=\!V_{TAPS}+ V_{QTPS} +V_S$, 
where~ $V_{TAPS}$ and $V_{QTPS}$ are given in Eqs. (1b) and (3), respectively\cite{prl_2}, 
and  $V_S \!=\!R_SI$. 
$R_S$ is a series resistance which includes the
contribution of the pads and other ohmic-like contributions such as
proximity effect of the normal pads on the SC wire\cite{boogaard}.
The fitting parameters are $R_{TAPS}$, $R_{QTPS}$ and $R_S$\cite{prl_2}, 
while
$a_{GIO}(=\!1.2)$ and $T_c(H)$, which enters through $\tau_{\scriptscriptstyle GL}$, 
are taken from the previous fits 
(Fig. 4(a)). At all temperatures and fields, the fitting curves reproduce the data extremely well, 
as shown in red in Figs. 2(a)-(b) and 3\cite{prl_heat}.  
At higher $T$ ($T/T_c\gtrsim 0.7$) the contribution from QTPS is negligible compared to TAPS.
At low $T$ ($T/T_c$ $\lesssim 0.5$) TAPS is expected to be exponentially suppressed (Eq. (1a)) 
and QTPS should dominate; this expectation is confirmed by the fits 
which yielded a negligible $R_{TAPS}$ compared to $R_{QTPS}$\cite{prl_9}.

\begin{figure}
\includegraphics[clip,width=0.45\textwidth]{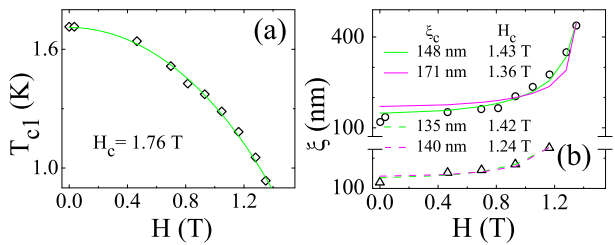}
\caption{(Color) 
GIO($\equiv$GIO+LAMH) 
fitting parameters to the linear resistance curves, including those 
shown in Fig.~2(c)-(d), plotted versus $H$.
(a) $T_{c1}$ ($\pmb{\pmb{\boldsymbol{\diamond}}}$) and $T_{c2}$ (not shown) 
fit well to the pair breaking perturbation theory\cite{TinkhamBook} (green).
(b) $\xi$  ($\pmb{\pmb{\boldsymbol{\circ}}}$) is fitted to
both $\xi\!=\!\xi _c/\sqrt[4]{1-(H/H_c)^2}$ (magenta) ($\Delta\!\propto\!\sqrt{1-(H/H_c)^2}$
 and $\xi\!\propto\!1/\sqrt{\Delta}$\cite{TinkhamBook}), 
and to an ad hoc expression $\xi \!=\!\xi _c/\sqrt{1-(H/H_c)^2}$ (green);
$\xi_c\!\equiv\!\xi (H\!=\!0)$ and ~$H_c$ is the critical field.
$\xi$ is better described by the ad hoc expression.
 $\xi$ ($\pmb{\pmb{\boldsymbol{\bigtriangleup}}}$) derived from the fits to $R_{QPS}$, 
including those in Fig.~2.
(e)-(f), is fitted to theory as before (dashed lines)
yielding   $\xi _c\approx 135~nm$, within 5\% of the
 value in Table~1.
}
\end{figure}

To further substantiate the importance
of QTPS at low $T$ while at the same time rule out TAPS as a significant
cause of phase slips, we
directly compare fits of the nonlinear V-I curves 
using $V=V_{QTPS}+V_S$ and $V=V_{TAPS}+V_S$.
The fitting parameters for QTPS are $R_{QTPS}$, $R_s$, and for TAPS they are 
$R_{TAPS}$, $R_S$.
The curves are presented in Figs. 3(a)-(c) 
after subtracting the linear background for several $H > 0$ T.
Although the two expressions have the same number of parameters,
the QTPS fits (red) are of good quality while the TAPS fits (blue) 
are evidently much poorer.
Varying $T$ by $\pm 50\%$ in the TAPS fits still failed to
reproduce the data, as shown in Fig. 3(a) for $T=0.59$ K and $H=0.465$ T.  
Comparing the $\chi$-squares of the fits that include QTPS versus TAPS alone
using the statistical  F-test yielded strong support for the importance 
of quantum phase slips with a confidence level $>$99.99\%\cite{prl_8}.

The $R_{QPS}=R_{QTPS}+R_{TAPS}$ extracted from the nonlinear V-I fits
enables to exclude contributions to residual resistance that are
irrelevant to phase slips, providing a means to check full
consistency. 
In Figs. 2(e)-(f) we 
plot $R_{QPS}$ ($\tud$ symbols) for $T\lesssim T_c$ and replot the 
linear resistance through the N-S transition for comparison; we then 
 refit the linear resistance expressions
(GIO+LAMH, Eqs. (1a) + (2)) to these data, keeping $a_{GIO}=1.2$ unaltered 
while varying
$T_c$ and $\xi$.  The discrete  $R_{QPS}$ data points versus $T$ no longer
exhibit a shoulder feature, and  only one $T_c$ was needed.
Satisfactory agreement is achieved 
as depicted in Figs. 2(e)-(f) (red).
Similar agreement comparable to the $H=0.465$ T trace is obtained
at higher $H =0.7$ T, 0.93 T, etc.
$T_c$ was essentially unchanged
compared to $T_{c1}$ in Fig.~4(a) but $\xi$  (Fig. 4(b), $\pmb{\pmb{\boldsymbol{\bigtriangleup}}}$)
was reduced at higher $H$, compared to the data obtained previously
(Fig. 4(b), $\pmb{\pmb{\boldsymbol{\circ}}}$).  
Fitting to the new values 
yielded $\xi _c\equiv \xi (H\!=\!0)\approx 135~nm$, within 5\% of the
calculated value in Table 1. As a final consistency check, 
analysis of the wider superconducting sample s1\cite{altomare_apl} did not produce any sign of
QTPS for $H\lessapprox 0.9$~T. This is sensible because of the larger
free-energy barrier.

The consistency achieved in our analysis of the nonlinear V-I curves and
residual resistance, 
supported by the quantitative statistical F-test, demonstrates the
importance of quantum phase slips.  This helps rule out other 
scenarios, e.g. TAPS alone or weak links. 
Furthermore, our results establish that
the transport properties of 1D superconducting nanowires  at temperatures 
much below $T_c$ are determined
primarily by the macroscopic quantum tunneling of phase slips.
Our findings pave the way to the study of newly predicted quantum 
phase transitions in metallic nanowires\cite{sachdev}.

F.A. thanks M.E. Rizza for her support.
Work 
supported by NSF DMR-0135931 and DMR-0401648.


\begin{thebibliography}{27}
\bibitem[{prl({\natexlab{a}})}]{prl_6} \bibinfo{note}{Linear
    resistance measurements were performed with a lockin-amplifier at
    23 Hz. Room temperature filters and low temperature RC filters,
    formed naturally from the sample lead resistances ($\approx$2-20
    $k\Omega$) and capacitances to ground ($\approx$0.5~nF), helped
    reduce noise to the nanowires.}
\bibitem[{prl({\natexlab{c}})}]{prl_2} \bibinfo{note}{{To distinguish
      the fitting parameters from the theoretical values, we refer to
      $R_{TAPS}$ ($R_{QTPS}$) as the TAPS (QTPS) resistance extracted
      from fits to the V-I, and to $R_{LAMH}$ ($R_{GIO}$) as the
      calculated values for TAPS (QTPS)}}.
\bibitem{GIO} In the following, when referring to the GIO theory,  
we will implicitly include the LAMH contribution as well.
\bibitem[{prl({\natexlab{b}})}]{prl_9} \bibinfo{note}{For the data
    central to QTPS shown in Figs. 2(f) and Fig. 3, there were a
    total of 21 fitting parameters for 1106 data points in the 10
    traces, or 53 points per parameter.}
\end{thebibliography}
\end{document}


\title{Erratum: Evidence for Macroscopic Quantum Tunneling of Phase Slips in Long One-Dimensional Superconducting Al Wires}

\author{Fabio Altomare} \author{Albert M. Chang} \affiliation{Physics
  Department, Duke University, Durham, NC 27708}\altaffiliation[Current address.]{} \affiliation{Physics
  Department, Purdue University, West Lafayette, IN 47906 }

\author{Michael R. Melloch} \affiliation{School of Electrical and
  Computer Engineering, Purdue University, West Lafayette, IN 47906}

\author{Yuguang Hong} \author{Charles W. Tu} \affiliation{Department
  of Computer and Electrical Engineering, UCSD, La Jolla, CA 92093}

\date{\today}

\pacs{74.78.Na, 74.25.Fy, 74.25.Ha, 74.40.+k}
\keywords{nanowire, stencil, aluminum, superconductivity, 
one-dimensional, macroscopic quantum tunneling, phase slip}

\maketitle
The definition of $\tau_{\scriptscriptstyle GL}$ should read  
$\tau _{\scriptscriptstyle GL}\!=\!(\pi/8)[\hbar/k_B(T_c-T)]$.

An inadvertent error in fitting to the magneto-resistance of the Al wires 
led to an underestimate of the wire widths and cross sections \cite{prl_5}.
  The revised width/cross section for the Al wires 
s1 (10 $\mu$m long) and s2 (100 $\mu$m long) are respectively: 
11.4 nm /130 nm$^2$ and 7.5 nm/58 nm$^2$ roughly an increase of 40 \% 
in width and factor of 2 in cross section.  The wire widths have 
an error of $\pm $ 10 \%, and the cross section error is $\pm $  20 \%.  
The revised sample parameters are listed in the revised Table I.  Note
that the wire length, L, and normal state resistance, $R_N$, remain
unchanged.

\begin{table}
\begin{tabular}{ccccccccc}\hline
   & L & w& t& $R_N$ & $\rho$& $l_e$ & $\xi$ & $J_c$ (0.35 K)\\ 
   & $\mu$m & nm & nm & $k\Omega$ & $\mu \Omega cm$ & nm & nm & $10^5
 A/cm^2$\\
\hline s1 & 10 & 11.4 & 11.4 & 8.3 & 10.7 & 3.7 & 65 &5.8  \\
 s2 &100 &7.5 &7.7 & 86.4 &5.0& 7 & 90 & 7.2 \\
\hline
\end{tabular}
  \caption{Parameters for sample s1 and s2. The parameters L (length),
  w (width), t (thickness), $R_N$ (normal state resistance), $\rho$
  (resistivity), $l_e$ (mean free path), $\xi$ (GL coherence length),
  and $J_c$ (critical current density), for the two samples s1 and
  s2. w, t, $\rho$ and $l_e$ have been deduced following the analysis
  in Altomare et al.\cite{altomare_apl} based on magnetoresistance
  measurements. $\xi$ was calculated using $\xi \!=\!0.85\sqrt{\xi _0
  l_e}$ where $\xi_0 \!=\!1600~nm$ is the GL superconducting coherence
  length.}
  \label{tab:tab1}
\end{table}

The dependence of the free-energy barrier $\Delta F$ on the cross
sectional area $A$ is contained in the quantity $L/R_N$\cite{Lmb+01}, 
removing any explicit dependence on $A$.  Thus, the revised quantities, 
including $A$, are not used in the main analysis, and do not affect the
 main conclusion of the paper.  The only notable effect is that the
 agreement between the fitted coherence length $\xi_c$ and the calculated 
$\xi$ now occurs at the boundary of the error bars.  Note that the error on
$\xi_c$ is ± 15\%.  The fitted values for $\Delta F(H=0,T=0)$ and 
$\xi_c(H=0,T=0)$ (135 nm), combined with the revised $A$,
 yield a thermodynamic critical field $H_{th}(H=0,T=0) \approx 0.0096$ T 
(96 Oersted), close to the known value of 0.0105 T (105 Oersted) 
for bulk aluminum.